\documentclass[twocolumn,showpacs,preprintnumbers,aps,amsmath,amssymb]{revtex4}

\usepackage{graphicx}
\usepackage{dcolumn}
\usepackage{bm}
\usepackage{epsfig}

\begin{document}


\title{Strain field of soft modes in glasses}

\author{U. Buchenau}
\affiliation{%
Forschungszentrum J\"ulich GmbH, J\"ulich Centre for Neutron Science (JCNS-1) and Institute for Complex Systems (ICS-1),  52425 J\"ulich, Germany}%

\date{February 6, 2021}

\begin{abstract}
The strain field surrounding the center of low frequency vibrational modes is analyzed for numerically created binary glasses with a $1/r^{10}$ repulsive interatomic potential. Outside the unstable inner core of five to twenty atoms, one finds a mixture of a motion similar to the string motion in the core with the strain field of three oscillating elastic dipoles in the center. The additional outside string motion contributes more to the stabilization of the core than the strain field, but the strain field dominates at long distances, in agreement with recent numerical findings. The small restoring force of the outside string motion places its average frequency close to the boson peak. The average creation energy of a soft mode in this binary glass is about 2.5 times the thermal energy at the freezing temperature. Scaling the soft potential parameters of the numerical modes to metallic glasses, one finds quantitative agreement with measurements of the sound absorption by tunneling states at low temperatures and by the excess modes at the boson peak.
\end{abstract}

\maketitle

\section{Introduction}

Low frequency localized vibrations in glasses are traditionally associated with the tunneling states \cite{philtun} and other low temperature or low frequency glass anomalies like the plateau in the thermal conductivity and the boson peak  \cite{bggprs,olig,parshin,ramos,corei,schiruo,buscho}. At present, they are heavily studied numerically \cite{le1,gale1,anael,manning,le2,corein,mizuno,wang,le3,coslovich,ikeda2,ikeda3,le4,le5,le6}, mostly in binary glasses, the numerical counterpart of binary metallic glasses \cite{johnson}.

An essential difference between real and numerical glasses is the finite size of the simulation cell, excluding long wavelength sound waves no longer fitting into the periodic boundary conditions \cite{taraskin}. The finite-size effects have been intensely studied \cite{mizu,caroli}. The main effect which remains even at very large system sizes is the delocalization of the modes with frequencies higher than the one of the lowest frequency sound wave of the simulation cell.

On the other hand, the number $Ne$ of atoms in the localized modes, calculated from the participation ratio $e$ (the inverse of the sum over the square of the squared eigenvector for all $N$ atoms in the simulation cell), is not $N$-dependent for large enough $N$. This shows that one can study the physical properties of these localized modes in moderate cell sizes without fearing a fundamental difference with their properties in extended samples.

Another way to deal with this uncertainty is to look at the third and fourth derivatives of the potential \cite{gale1}. One can not only define harmonic eigenmodes in the usual way by considering the eigenvalues of the second derivatives of the potential with respect to the atomic coordinates, but also third order and quartic eigenmodes by an appropriate extension. The modes obtained in this way are not identical with the harmonic ones, but have very similar properties, though the quartic modes are better localized than the harmonic ones in the coexistence region with the sound waves \cite{gale1,bule}.   

The core of the modes shows a string motion of the atoms with the largest amplitude (a nearly parallel motion of the two neighbors of an atom along the direction of its displacement) \cite{olig} and is unstable \cite{corei,buscho,corein}. The boundary of the unstable core is defined by the atoms which begin to have a positive contribution to the restoring force of the mode \cite{corein}. Its saddle point configuration in energy is stabilized by positive restoring forces from the surroundings. 

A very recent study \cite{le6} concludes that the long range distortion field is describable in terms of the elastic Greens function of an isotropic elastic medium, assuming elastic dipoles situated in the core center. At first sight, the result seems to support the earlier hypothesis \cite{buscho} that the elastic distortion around holds the core on its saddle point. As will be seen here, this is only partially true.

The present paper analyzes the balance between inner and outer forces on numerical soft modes, in order to assess its significance for our understanding of glasses and the glass transition. The second aim of the paper is to establish a quantitative link between numerical data and experiments in real glasses in terms of the soft potential model \cite{bggprs,parshin,ramos}. 

The analysis itself is presented in Section II. Its consequences for experiments in real glasses are derived in Section III. Section IV discusses the possible implications of the frozen saddle points for the boson peak and the glass transition, the analogies and differences between binary numerical and metallic glasses, and ends by summarizing the paper.  

\section{Analysis}

\subsection{Soft modes}

The analysis was done on twenty soft modes calculated by the Amsterdam group \cite{le1,gale1,le2,le3,le4,le5,le6} for a 50:50 binary mixture of atoms with the 3DIPL potential between two small atoms
\begin{equation}	\frac{E(r)}{a_0^{10}}=E_0\left(\frac{1}{r^{10}}-\frac{56}{r_c^{10}}+\frac{140r^2}{r_c^{12}}-\frac{120r^4}{r_c^{14}}+\frac{35r^6}{r_c^{16}}\right)
\end{equation}
with a cutoff at $r_c=1.48a_0$. $a_0$ is the length unit, $E_0$ the energy unit of the system. The energy and its first, second and third derivative with respect to $r$ are zero at the cutoff.

Between small and large atoms, one has the same potential with $a_0$ replaced by 1.18$a_0$, and between large atoms by 1.4$a_0$, adapting $r_c$ accordingly.
  
In terms of the potential energy $E_0$, the modes were frozen in at a temperature $T_g$ with $k_BT_g=0.5E_0$. In the freezing, the atomic volume $V_a$ was held constant at $1.22a_0^3$, where $a_0$ is the length unit (87 percent of the average nearest neighbor distance). The mass unit $M$ is 1 for both kinds of atoms. The frozen glasses have a bulk modulus $B$ of 56 $E_0/a_0^3$ and a shear modulus $G$ of 14 $E_0/a_0^3$, leading to a transverse sound velocity of 4.12 $E_0^{1/2}M^{-1/2}$, a longitudinal sound velocity of 9.52 $E_0^{1/2}M^{-1/2}$, and a Debye frequency $\omega_D$ of 17.2 $E_0^{1/2}M^{-1/2}/a_0$. Ten of the modes were normal modes in a cell of 16000 atoms, the other ten were quartic modes \cite{gale1} in a cell of 128000 atoms.

To find the center of the mode, one first looks for atom 1 with the largest eigenvector amplitude $\bm{e}_1$. One takes a sphere of radius 3.5$a_0$ (about three nearest neighbor distances) around this atom, containing a number $N$ of about hundred and forty atoms, and calculates a new center $\bm{R}_0$ with
\begin{equation}
	\bm{R}_0=\sum_{i=1}^N\bm{r}_i\bm{e}_i^2/\sum_i^N\bm{e}_i^2,
\end{equation}
where $\bm{r}_i$ and $\bm{e}_i$ are the position vectors and the eigenvector amplitudes of the inner core atoms, respectively.

Iterating this procedure around the new center three times, one finds a mode center which does not change any longer, and which is always closer than $a_0$ to the starting point. The core defined in this way contains about sixty percent of the normalized squared eigenvector for most modes.

\subsection{Core energy calculation}

For a given core radius $r_c$ around this center, one can calculate the energy of the core atoms as a function of the mode coordinate by summing their interaction energy with the other atoms in the core and with those atoms in the outer shell which are still within the interaction range (for the 3DIPL potential about 2$a_0$).

\begin{figure}[t]
\hspace{-0cm} \vspace{0cm} \epsfig{file=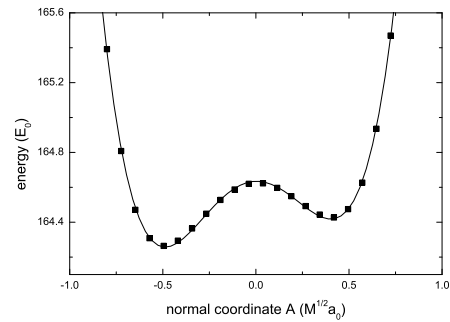,width=6 cm,angle=0} \vspace{0cm} \caption{Potential energy for the core of mode m3 with radius $r_c=1.9a_0$. The line is a soft potential fit in terms of eq. (\ref{ea}).}
\end{figure}

\begin{figure}[b]
\hspace{-0cm} \vspace{0cm} \epsfig{file=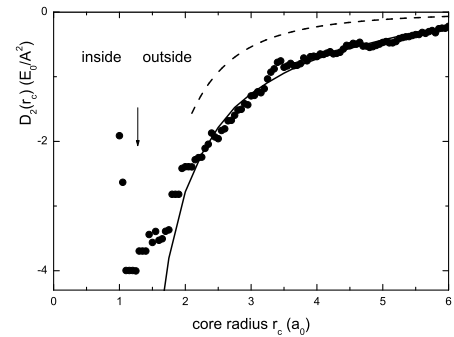,width=6 cm,angle=0} \vspace{0cm} \caption{Negative restoring force constant $D_2$ as a function of the core radius $r_c$ for the same mode m3. Dashed line is $D_2(r_c)$ for only elastic outside restoring forces, continuous line elastic plus additional restoring forces from random modes outside the core (see text). The arrow marks the point where the atoms begin to contribute positive restoring forces, taken to be the inner core radius.}
\end{figure}

Fig. 1 shows this energy dependence for the normal mode m3 with the core radius 1.9 $a_0$, containing 21 atoms. The instability of the core is clearly seen. The dependence is well fitted in terms of the soft potential function \cite{bggprs,parshin,ramos}, a polynomial up to fourth order in $A$  
\begin{equation}\label{ea}
	E(A)=E(0)+D_1A+D_2A^2+D_3A^3+D_4 A^4,
\end{equation}
where $A$ is the normal coordinate, defined by the kinetic energy $E_{kin}=\dot{A}^2/2$ of the mode.

With increasing $r_c$, $D_4$ saturates to a constant value, for most of the modes below $r_c=2.5a_0$. $D_2(r_c)$ is negative, but increases with increasing $r_c$. This is shown in Fig. 2 for the same mode m3 as in Fig. 1. The arrow marks the boundary between inside and outside defined by the crossover to positive restoring forces, which lies considerably lower than the saturation value for $D_4$.

The simplest explanation for the increase of $D_2(r_c)$ is an elastic distortion of the surrounding elastic medium \cite{buscho,le6} which is proportional to the normal coordinate $A$ and which provides the positive restoring forces holding the unstable core on its saddle point in energy. It has recently been shown \cite{asyth4} that one can understand the surprisingly strong temperature dependence of the boson peak frequency in the glass phases of silica \cite{wischi} and polycarbonate \cite{schoenfeld} in terms of this concept, postulating an unstable core also for the boson peak modes.

\subsection{Elastic strains}

For the numerical modes, one can check whether this is indeed the correct explanation by fitting the mode displacements within an outer shell with shell radius $r_s$ and the thickness $d$ by an elastic dipole tensor proportional to $A$ in the core center \cite{leibfried,newfrench,le6}. Such an elastic dipole tensor gives rise to an atomic displacement decreasing with $1/r_s^2$, leading to a mean square eigenvector $<e^2>_{shell}$ for the atoms in the shell decreasing proportional to $1/r_s^4$. The elastic displacement vector $\bm{s}$ can be calculated from the elastic Greens function
\begin{equation}
	G_{ij}(\bm{r})=\frac{1}{8\pi G}\left(\frac{\delta_{ij}(1+f)}{r}+\frac{x_ix_j(1-f)}{r^3}\right)
\end{equation}
where
\begin{equation}
	f=\frac{3G}{3B+4G}=\frac{v_t^2}{v_l^2}.
\end{equation}
$\delta_{ij}$ is the Kronecker symbol, $x_i$ and $x_j$ are cartesian components of the position vector $\bm{r}$. The displacement vector $\bm{s}$ is given by
\begin{equation}
	\bm{s}_i(\bm{r})=-\sum_{j,k=1}^3\frac{\partial G_{ij}}{\partial x_k}P_{jk},
\end{equation}
where the $P_{jk}$ are the six components of the symmetric elastic dipole tensor in the center of the core. $P_{jk}$ is in principle the first moment of a distribution of $N$ forces around a point defect \cite{leibfried,newfrench}
\begin{equation}
	P_{jk}=\sum_{i=1}^NF_{ij}r_{ik},
\end{equation}
with $F_{ij}$ the component of the force $i$ in $j$-direction, and $r_{ik}$ the component of the application point of the force in $k$-direction. Here, however, the six $P_{jk}$-values are simply treated as fit variables to determine the long-range elastic distortion around the core center. 

But if one does these fits, one finds that the outside elastic response is only part of the explanation, the smaller part as far as the stabilization of the inner core on its saddle point in energy is concerned. The elastic dipole tensor displacements dominate at large distances from the core, as established by the recent numerical study \cite{le6}, but at smaller distances they are overshadowed by another more random kind of motion. This is demonstrated in Fig. 3, which shows the product $<e^2>_{shell}r_s^4$ for consecutive shells of mode m3 with $d=0.5$. The dashed line is the fitted dipole tensor contribution. 

\begin{figure}[t]
\hspace{-0cm} \vspace{0cm} \epsfig{file=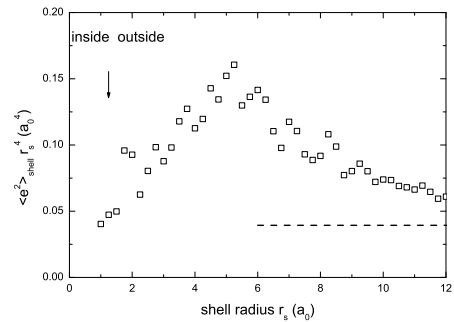,width=6 cm,angle=0} \vspace{0cm} \caption{Product of mean square displacement $<e^2>_{shell}$ with the fourth power of the shell radius $r_s$ for consecutive shells of mode m3 with a shell thickness of $0.5\ a_0$, showing the excess mean square displacement over the elastic one (the dashed line).}
\end{figure}

The elastic part of the positive restoring force contribution can be calculated with the elastic Greens function \cite{leibfried,newfrench}. For the simplest case of an expansion $\delta V/V$ of the core volume, one finds the outside energy $2GV(\delta V/V)^2/3$ of the shoving model \cite{shov}, for a pure shear or a single dipole distortion of the core an energy which is a factor 1.22 larger for the same sum of squares of the $P_{jk}$ (the factor 1.22 is calculated for the Poisson ratio 0.384 of the 3DIPL potential, close to the average Poisson ratio of 31 metallic glasses \cite{johnson}).  If this elastic force were the only positive restoring force from the outside (and if the frequency of the mode were exactly zero), $D_2(r_c)$ should follow the dashed line in Fig. 2. 

The dipole tensor fit is reliable, because it yields the same six tensor elements within experimental error for different shell diameters $r_s$, if they are larger than $6a_0$ and still small compared to the distance of the dipole tensor in the neighboring periodic cell (26.9 $a_0$ for the cell with 16000 atoms). By a suitable rotation of the coordinate system, one can diagonalize the tensor and describe it by its three diagonal elements $P_{11},P_{22}$ and $P_{33}$. In the rotation, the new $x$-axis was chosen in the direction of largest expansion and the new $y$-axis along the direction of the strongest contraction. Looking at the tabulated values in Table I, one realizes that the elastic distortion component of the motion is essentially a shear, because the sum of the three diagonal elements is small for all investigated modes.

The very recent numerical investigation \cite{le6} of the strain field around the core of the modes comes to the conclusion that it is essentially a simple planar shear. The data in Table I agree with this conclusion, because the planar shear dipole $(P_{11}-P_{22})/\sqrt{2}$ is much larger than the independent second shear dipole $(P_{11}+P_{22}-2P_{33})/\sqrt{6}$ of a diagonal elastic dipole tensor.  

\subsection{Additional outside motion}

One can subtract the elastic dipole displacement from the eigenvector of the atoms outside the unstable core to study the more random motion. After subtracting, one finds that the radial component of the displacement of the outside atoms on the average still remains closer to $1/\sqrt{2}$ of the total than to the $1/\sqrt{3}$ of a completely random motion. Also, neighboring atoms still tend to move together, especially when they move along the line connecting them. Thus the more random component of the outside motion shows the string character first observed around the atom with the largest displacement in the inner core \cite{olig}, suggesting that one has a similar kind of motion inside and outside, with the difference that the average restoring force is negative inside and positive outside.

To describe the measured $D_2(r_c)$ in Fig. 2, it is natural to assume that the energy of the more random component is not proportional to the derivative of the displacement (as in the elastic dipole component), but rather proportional to the mean square displacement $<u^2>_r$ of the more random component itself, with the (adaptable) proportionality factor $f_r$. Then its contribution to $D_2(r_c)$ is given by
\begin{equation}
	D_{2r}(r_c)=-f_r\sum_{r_i>r_c}\bm{e}_{ir}^2 
\end{equation}
where $\bm{e}_{ir}$ is the random component of the eigenvector of atom $i$ at a distance $r_i$ from the mode center, obtained by subtracting the fitted elastic component. Adding this to the dashed line in Fig. 2 with $f_r=1.6$, one gets the continuous line, which describes $D_2(r_c)$ very well, at least as long as one does not approach the inner core below $r_c\approx1.5\ a_0$. One concludes that the larger part of the stabilizing outside restoring force is not due to the elastic distortion, but rather to the interaction with other string-like modes, proportional to their mean square displacement.

Naturally, one is dealing here with orthogonal modes which do not have a bilinear coupling between them. But this does not exclude bilinear interactions between different soft spots which then enter the eigenvector of the resulting orthogonal modes and show up in the present analysis.

$f_r$ is an average restoring force of the more random modes surrounding the core, to be compared with the restoring force $\omega_D^2/2=147.9$ of the Debye frequency. In Table I, it varies between 1.6 and 5.6 from mode to mode, showing that these modes come from a broad frequency band between one tenth and one fifth of the Debye frequency, in metallic glasses the frequency region from 2.5 to 5 meV at or slightly above the boson peak \cite{suck,scho1,scho2}. For the modes in Table I, they always supply at least two thirds of the stabilizing restoring forces. 

\subsection{Bilinear coupling}

On the other hand, the elastic distortion dominates at long distances, and so determines the bilinear coupling of the mode to an external strain
\begin{equation}
	E_{bilinear}=P_{11}\epsilon_{11}A+P_{22}\epsilon_{22}A+P_{33}\epsilon_{33}A.
\end{equation}

Of the five possible shears $\epsilon_t$ for a transverse sound wave, only two interact with the diagonal dipole tensor, namely the one with $\epsilon_{11}=-\epsilon_{22}=\epsilon_t/\sqrt{2}$ and the other with $2\epsilon_{11}=2\epsilon_{22}=-\epsilon_{33}=\epsilon_t/\sqrt{3/2}$. Averaging the squares, one finds an average bilinear transverse sound wave coupling $\lambda\epsilon_tA$ with
\begin{equation}
	\lambda_t^2=\frac{1}{15}(P_{11}^2+P_{22}^2+P_{33}^2-P_{11}P_{22}-P_{22}P_{33}-P_{33}P_{11}).
\end{equation}

The longitudinal strain $\epsilon_l$ is the combination of an expansion $\epsilon_l$ with the shear strain $4\epsilon_l/3$, so in the bilinear longitudinal sound wave coupling $\lambda_l\epsilon_lA$
\begin{equation}
	\lambda_l^2=(P_{11}+P_{22}+P_{33})^2+\frac{16}{9}\lambda_t^2.
\end{equation}

\begin{table}[htbp]
	\centering
		\begin{tabular}{|c|c|c|c|c|c|c|c|c|}
\hline
mode                      &  $f_r$  &$P_{11}$&$P_{22}$&$P_{33}$& $E_s$  &$\Delta A$ & $Ne$ & $D_4$    \\
\hline
                          &$E_0/A^2$& $E_0/A$&$E_0/A$ &$E_0/A$ &$E_0$   & $A$       &      &$E_0/A^4$ \\
\hline
 m0                       &  4.3    &   41   & -55    &   7    & 1.7    &   1.0     & 17.6 &    9.3   \\
 m1                       &  2.4    &   52   & -49    &   2    & 0.5    &   0.9     & 65.8 &    4.0   \\
 m2                       &  3.1    &   39   & -54    &   9    & 0.9    &   1.0     & 16.6 &   20.5   \\
 m3                       &  1.6    &   48   & -53    &  19    & 0.9    &   1.0     & 35.1 &    7.3   \\
 m4                       &  4.9    &   45   & -46    &  11    & 0.7    &   0.9     & 12.5 &   16.5   \\
 m5                       &  3.5    &   22   & -51    &  17    & 1.8    &   1.2     & 34.3 &    7.9   \\
 m6                       &  4.8    &   45   & -36    & -20    & 0.4    &   0.7     & 11.1 &   21.1   \\
 m7                       &  5.6    &   61   & -55    & -20    & 0.6    &   1.0     & 13.0 &   10.1   \\
 m8                       &  2.7    &   29   & -34    &  13    & 1.7    &   1.3     & 77.3 &    2.9   \\
 m9                       &  4.4    &   35   & -55    &  21    & 3.1    &   1.4     & 37.8 &    6.2   \\
\hline                                                                       
 q0                       &  5.1    &   46   & -23    &  -5    & 2.5    &   1.0     & 50.6 &    6.6   \\
 q1                       &  3.2    &   31   & -39    &   8    & 1.3    &   0.9     & 43.8 &    4.2   \\
 q2                       &  2.2    &   34   & -31    & -19    & 0.7    &   1.4     &  8.1 &   19.4   \\
 q3                       &  1.6    &   51   & -32    & -15    & 0.4    &   1.0     & 38.2 &    7.9   \\
 q4                       &  1.8    &   50   & -36    &  25    & 0.7    &   1.0     & 36.0 &    6.3   \\
 q5                       &         &   45   & -58    &  -3    &        &           & 99.5 &    2.6   \\
 q6                       &  2.4    &   46   & -66    &  13    & 0.8    &   1.2     & 30.4 &   10.9   \\
 q7                       &  4.9    &   40   & -33    & -10    & 2.3    &   1.0     & 21.5 &    6.1   \\
 q8                       &  4.6    &   40   & -33    & -10    & 1.6    &   1.2     & 28.2 &   10.3   \\
 q9                       &         &   92   & -12    &  30    &        &           & 21.4 &   575    \\ 
 \hline
 		\end{tabular}
	\caption{Results of the unstable core analysis for ten normal modes (m) and ten quartic modes (q). Note that $A$, the normal coordinate unit, is $M^{1/2}a_0$, so $f_r$ is a squared frequency.}
	\label{tab:rse1}
\end{table}
For the ten normal modes in Table I, one finds average values of 29.5 $E_0/A$ for $\lambda_l$ and 21 $E_0/A$ for $\lambda_t$. The ratio $\lambda_l/\lambda_t$ is 1.4, close to the value 4/3 for a pure shear distortion and only slightly smaller than the average ratio 1.6 found for tunneling states in a data collection \cite{berret}.

\subsection{Soft mode creation energy} 

The next step is to estimate the creation energy of the soft modes, allowing the core atoms to find their energy minimum for a given mode displacement of the outer atoms, by an appropriate variation of the core atom positions and an adaptation of the mode displacement of the outer atoms. In this way, one can find the minimum packing energy of the core atoms for the given density. The difference between this minimum packing energy and the energy at displacement zero provides the frozen saddle point energy $E_s$, the creation energy of the soft mode. The calculations were done for the core radius $r_c=2.6a_0$. The analysis was not done for mode q5, where $D_4$ saturates very late, at 4.4 $a_0$, and also not for mode q9, which had no unstable core and a very large $D_4$.

It turns out that this minimum of the packing energy occurs for a relatively large displacement of the outer atoms along the mode coordinate, about 1.5 times larger than the displacement for the minima in Fig. 1. About eighty percent of the atomic motion inside is a backward motion along the mode coordinate, bringing the inside mode diplacement down to the $\Delta A$ in Table I, about two thirds of the outside value. The accurate determination of the two minimal energies in both directions requires the subtraction of the linear potential term. The average creation energy is 1.25 $E_0$ (2.5 $k_BT_g$) for both the normal modes and the quartic modes. The value 2.5 $k_BT_g$ for the creation energy $E_s$ of the soft modes explains the Kohlrausch stretching exponent close to 1/2 according to a recent theory of the highly viscous flow \cite{ac}, and should be more or less the same in all glass formers in order to explain the universal density of tunneling states \cite{philtun}, according to a recent explanation of Kauzmann's paradox \cite{kauz}.

The creation energy of 2.5 $k_BT_g$ of the soft modes explains the strong reduction in their number density in glasses \cite{wang} cooled to low glass temperatures with the swap mechanism \cite{swap}. Obviously, the number density of soft modes in the glass increases with increasing freezing temperature of the undercooled liquid.

\subsection{Soft potential parameters}

Finally, Table I lists the two parameters usually characterizing the soft modes in the numerical world, the number $Ne$ of atoms in the mode, calculated from the participation ratio $e$ and the fourth order term $D_4$.

From the latter, one can calculate the crossover energy $W$ between tunneling states and soft vibrations of the soft potential model \cite{bggprs,parshin,ramos}. $W$ is the quantum mechanical zero point energy in the purely quartic potential, given by the equality of potential and kinetic confinement energy at the normal coordinate $A_0$
\begin{equation}\label{zp}
	D_4A_0^4=\frac{\hbar^2}{2A_0^2}\equiv W
\end{equation}
leading to
\begin{equation}\label{wa0}
	W=\frac{\hbar^{4/3}D_4^{1/3}}{2^{2/3}}\ \ \ A_0=\frac{\hbar^{1/3}}{2^{1/6}D_4^{1/6}}.
\end{equation}

Averaging $D_4^{1/3}$ over the ten normal modes in Table I, one arrives at $W=1.33$ in energy units $E_0^{1/3}(\hbar^2/Ma_0^2)^{2/3}$ and $A_0=0.785$ in normal coordinate units $M^{1/2}a_0(\hbar^2/(E_0Ma_0^2))^{1/6}$. Note that the averaging over $D_4^{1/3}$ reduces the scatter of $D_4$ on a logarithmic scale by a factor of three, making $W$ reasonably well defined.

\section{Comparison to metallic glasses}

\subsection{Tunneling states}

Unlike a numerical glass, a real glass has to face quantum mechanics at low temperatures, requiring the application of equs. (\ref{zp}) and (\ref{wa0}) to describe not only soft vibrations, but also tunneling states in terms of the soft potential model \cite{bggprs,parshin,ramos}, an extension of the tunneling model \cite{philtun}. In a metallic glass, one has the additional difficulty that the tunneling states do not only interact with the sound waves, but also with the conduction electrons \cite{golding}.

The metallic glass most heavily studied \cite{golding,ray,esqui} at low temperatures is Pd$_{78}$Si$_{16}$Cu$_6$, with an average atomic mass of 91.3 a. u. and an atomic volume of 0.046 nm$^{3}$, so $a_0=0.229$ nm. Thus the kinetical confinement energy unit $\hbar^2/(Ma_0^2)=8.64$ $\mu$eV. $E_0=0.208$ eV can be calculated from the shear modulus $G=31.8$ GPa \cite{johnson} and the known $a_0$. With these two values, eq. (\ref{wa0}) yields $W/k_B=3.85$ K, close to the value 3.2 K found in a recent soft potential evaluation \cite{abs} of the crossover from the tunneling plateau to classical relaxation in the experimental data.

The value $A_0$ calculated with these values for $E_0$ and $\hbar^2/(Ma_0^2)$ from the second part of eq. (\ref{wa0}) allows to determine the coupling constant $\Lambda_t=\lambda_tA_0$ of the soft potential model, $\Lambda_t=0.64$ eV. From this value, it is straightforward to calculate the phonon-coupling parameter $\gamma_t\approx\Lambda_t/\sqrt{3}=0.37$ eV of the tunneling model, in excellent agreement with the value 0.4 eV found in the pioneering tunneling model fit of high frequency sound absorption data \cite{golding} at low temperatures in this metallic glass. 

\subsection{Boson peak}

\begin{figure}[t]
\hspace{-0cm} \vspace{0cm} \epsfig{file=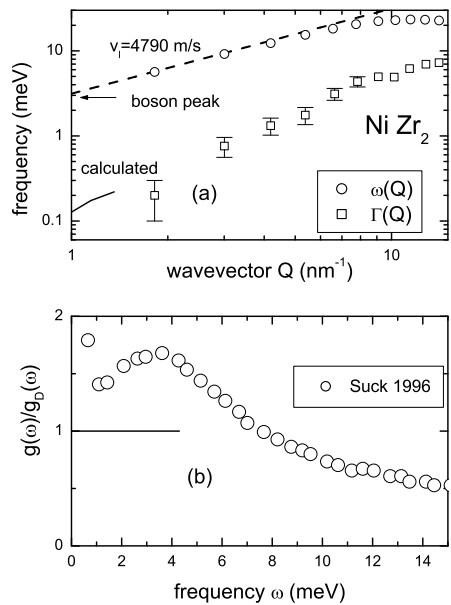,width=6 cm,angle=0} \vspace{0cm} \caption{(a) Frequency and damping of longitudinal sound waves measured by x-ray Brillouin scattering \cite{sco} in NiZr$_2$ compared to the calculation with the numerical coupling at the boson peak (b) Boson peak in NiZr$_2$ measured by inelastic neutron scattering \cite{suck} and normalized to the Debye density of states as described in the text.}
\end{figure}

One can further compare the numerical coupling constants to x-ray Brillouin data \cite{sco} in another metallic glass, NiZr$_2$. The damping of the longitudinal sound waves at the boson peak is given by
\begin{equation}
	\Gamma(\omega)=\frac{3\pi}{2}\frac{\lambda_l^2}{Mv_l^2}g_{exc}(\omega),
\end{equation}
where $g_{exc}(\omega)$ is the excess density of states over the Debye density of states $g_D(\omega)$ given by $3\omega^2/\omega_D^3$. $\Gamma(\omega)$ is the half width at half maximum of the Brillouin line.

The equation is valid as long as the first order perturbation approximation holds. It should hold up to the Ioffe-Regel limit $\Gamma(\omega)=\omega/\pi$, where sound waves cease to propagate. Fortunately, this limit is not yet reached in metallic glasses at the boson peak \cite{sco}, so the equation should be applicable there.

Fig. 4 (a) shows the comparison of measured x-ray Brillouin data \cite{sco} in the metallic glass NiZr$_2$ with this calculation. NiZr$_2$ has the longitudinal sound velocity 4790 m/s and the atomic volume 0.0188 nm$^3$. Its transverse sound velocity has not been measured, but one finds the ratio $v_t^2/v_l^2=3/16$ of the 3DIPL potential in good approximation in Cu$_{50}$Zr$_{50}$Al$_8$ \cite{li} and in Cu$_{46}$Zr$_{54}$ \cite{johnson}. With this ratio, one calculates a Debye frequency of 22.5 meV for NiZr$_2$. The ratio $\lambda_l^2/Mv_l^2$ is a squared frequency and scales with the square of the Debye frequency.

The boson peak of NiZr$_2$ has been measured by inelastic neutron scattering \cite{suck} and is shown in Fig. 4 (b). For its normalization to the Debye density of states, its height of 1.65 $g_D(\omega)$ was taken over from low temperature heat capacity measurements \cite{tian} in Cu$_{50}$Zr$_{50}$Al$_8$.

With this normalization, the short line on the left of Fig. 4 (a) was calculated (one cannot go beyond 4.3 meV, because the Debye density of states is bound to fall below its zero frequency value at higher frequency). Within the error bars of the Brillouin x-ray measurement \cite{sco}, the measured damping extrapolates to the calculated line, showing that the excess modes at the boson peak of this metallic glass have the same coupling to external strains as the numerical low frequency modes of the 3DIPL potential.

\section{Discussion and summary}

\subsection{Frozen saddle points}

The early recognition \cite{corei} of an unstable core of numerical soft modes has been recently corroborated in much more detail \cite{corein}, demonstrating that the small positive restoring force of these modes results from the overcompensation of the negative inside restoring force by positive restoring forces from the surroundings.

The present paper separates the positive outside restoring forces into a well understandable elastic distortion, describable by three elastic dipoles in the mode center, and a less well understandable bilinear interaction with neighboring soft spots.

The estimate of the creation energy of the frozen saddle point of Section II. F shows that the core freezes between two minima of the packing energy. The existence of a linear potential term demonstrates that the freezing can happen with equal probability at any point between the two minima. If the core freezes close to one minimum, one does not get a soft spot, defining the soft spot as a mode with an unstable core.

If there are soft and hard spots around, why then does one see only soft spots in the outside motion? Apart from their larger susceptibility, the answer must be that the mode is formed at the glass temperature, where its free energy profits from its mean square displacement \cite{se}. In fact, the soft modes seem to be the entropy carriers of the undercooled liquid \cite{kauz}. An intimate connection between soft modes and glass transition is also suggested by the finding of strings not only in the core of soft modes \cite{olig}, but also in the relaxational motion at the glass transition \cite{kob}. A recent treatment \cite{ac} of the highly viscous flow in terms of creation and annihilation of soft modes relates the Kohlrausch exponent $\beta$ of the viscous relaxation to the creation energy of the soft modes. The treatment is supported by the strong bilinear coupling between neighboring soft spots found in the present analysis.

The picture of neighboring soft spots with an unstable core undergoing a shear distortion with increasing displacement explains why one finds the final values for the three elastic dipoles $P_{11}$, $P_{22}$ and $P_{33}$ only for shell radii larger than 6 $a_0$, though the core radius defined by the onset of positive restoring forces lies below 1.5 $a_0$ for most of the modes in Table I. The reason is that surrounding soft spots between 1.5 and 6 $a_0$ still contribute to the elastic distortion at large distances, though their center does not coincide exactly with the one of the mode.

Another consequence of this consideration is that there is a large amount of elastic distortion which is not centered in the mode center, does not contribute to the three dipoles, but does contribute to the positive restoring forces of the mode. This raises the question whether for a given mode ultimately all positive force constants result from the elastic distortion, all negative force constants from soft spots, not only from the one in the center, but also from those in the surroundings. There is evidence \cite{asyth4} for this hypothesis from measurements of the anomalous temperature dependence of the boson peak frequency in two well-studied glasses \cite{wischi,schoenfeld}.  

In this context, it should be noted that the measurement of the boson peak in Fig. 4 (b) was done at two temperatures \cite{suck}, at 296 K and 400 K and shows no visible change. But then, the shear modulus of a similar metallic glass, Vit4, diminishes \cite{lind} only by 2.6 percent by heating from 296 K to 400 K. Also, the string motion inside might have the same Gr\"uneisen parameter as the sound waves, which would prevent any anomalous temperature dependence in the glass phase.

\subsection{Analogies and differences}

As pointed out in Section II. C, the Poisson ratio of the numerical 3DIPL glass is close to the one of 31 metallic glasses \cite{johnson}. But one should be aware that the binary numerical glasses are not the perfect counterpart to the metallic glasses. Their interatomic potential is much too steep for that; they are much closer to a frozen noble gas than to a metallic glass. Of course, a steep potential is a great numerical advantage, because it allows for a cutoff around two atomic distances and thus reduces the number of neighbors for which one must calculate the interatomic forces.

One can see the difference from the ratio $GV_a/k_BT_g$ ($V_a$ atomic volume), which for the 3DIPL potential is 34.1 and for the metallic glasses \cite{johnson} is 65.6 (not 17.6 as erroneously stated earlier \cite{buscho}). Apparently, the more harmonic interatomic potential in metals allows for a freezing into a glass with a much higher shear modulus.

But in spite of this marked difference, the analysis of twenty numerical 3DIPL modes provides a crossover energy $W$ between tunneling states and soft vibrations in good agreement with experimental data in metallic glasses. The same good agreement is found for the bilinear coupling to the sound waves, indicating that the nature of the soft modes in simple liquid is independent on the anharmonicity of the interatomic potential.

\subsection{Summary}

To summarize, the analysis of the positive restoring forces outside the inner unstable core of soft vibrational modes in numerically cooled binary glasses allows to separate a smaller part due to the elastic distortion and a larger part due to the interaction with other soft spots outside the core. The smaller elastic part determines the bilinear coupling of the mode to the sound waves. Scaling the results with the shear modulus, the atomic mass and the atomic volume, one gets quantitative agreement with measured low temperature sound absorption and x-ray Brillouin data in metallic glasses within the soft potential model.

The data support the hypothesis that the positive restoring forces stabilizing the unstable core on its saddle point in energy are ultimately due to elastic distortions, either mediated by surrounding soft spots or directly due to the elastic distortion of the core itself. 

Thanks are due to Edan Lerner for supplying the twenty soft modes, to Edan Lerner and Eran Bouchbinder for enlightening discussions and helpful suggestions.


\begin{thebibliography}{99}
\bibitem{philtun} W. A. Phillips, Rep. Prog. Phys. {\bf 50}, 1657 (1987)
\bibitem{bggprs} U. Buchenau, Yu. M. Galperin, V. L. Gurevich, D. A. Parshin, M. A. Ramos and H. R. Schober, Phys. Rev. B {\bf 46}, 2798 (1992)
\bibitem{olig} H. R. Schober, C. Oligschleger, and B. B. Laird, J. Non-Cryst. Solids {\bf 156-158}, 965 (1993)
\bibitem{parshin} D. A. Parshin, Phys. Solid State {\bf 36}, 991 (1994)
\bibitem{ramos} M. A. Ramos and U. Buchenau, Phys. Rev. B {\bf 55}, 5749 (1997) 
\bibitem{corei} V. A. Luchnikov, N. N. Medvedev, Yu. J. Naberukhin, and H. R. Schober, Phys. Rev. B {\bf 62}, 3181 (2000)
\bibitem{schiruo} W. Schirmacher, G. Ruocco, and T. Scopigno, Phys. Rev. Lett. {\bf 98}, 025501 (2007)
\bibitem{buscho} U. Buchenau and H. R. Schober, Phil. Mag. {\bf 88}, 3885 (2008)
\bibitem{le1} E. Lerner, G. D\"uring, and E. Bouchbinder, Phys. Rev. Lett. {\bf 117}, 035501 (2016)
\bibitem{gale1} L. Gartner and E. Lerner, SciPost {\bf 1}, 016 (2016)
\bibitem{anael} S. Gelin, H. Tanaka, and A. Lemaitre, Nat. Materials {\bf 15}, 1177 (2016)
\bibitem{manning} S. Wijtmans and M. L. Manning, Soft Matter {\bf 13}, 5649 (2017)
\bibitem{le2} G. Kapteijns, E. Bouchbinder, and E. Lerner, Phys. Rev. Lett. {\bf 121}, 055501 (2018)
\bibitem{corein} M. Shimada, H. Mizuno, M. Wyart, and A. Ikeda, Phys. Rev. E {\bf 98}, 060901 (2018)
\bibitem{mizuno} H. Mizuno and A. Ikeda, Phys. Rev. E {\bf 98}, 062612 (2018)
\bibitem{wang} L. Wang, A. Ninarello, P. Guan, L. Berthier, G. Szamel, and E. Flenner, Nat. Commun. {\bf 10}, 26 (2019)
\bibitem{coslovich} D. Coslovich, A. Ninarello, and L. Berthier, SciPost Phys. {\bf 7}, 077 (2019)
\bibitem{le3} G. Kapteijns, D. Richard, and E. Lerner, Phys. Rev. E {\bf 101}, 032130 (2020)
\bibitem{ikeda2} H. Mizuno, H. Tong, A. Ikeda, and S. Mossa, J. Chem. Phys. {\bf 153}, 154501 (2020)
\bibitem{ikeda3} H. Mizuno, M. Shimada, and A. Ikeda, Phys. Rev. Research {\bf 2}, 013215 (2020)
\bibitem{le4} C. Rainone, E. Bouchbinder, and E. Lerner, J. Chem. Phys. {\bf 152}, 194503 (2020)
\bibitem{le5} D. Richard, K. Gonzalez-Lopez, G. Kapteijns, R. Pater, T. Vaknin, E. Bouchbinder, and E. Lerner, Phys. Rev. Lett. {\bf 125}, 085502 (2020)
\bibitem{le6} A. Moriel, Y. Lubomirsky, E. Lerner, and E. Bouchbinder, Phys. Rev. E {\bf 102}, 033008 (2020) 
\bibitem{johnson} W.L. Johnson and K. Samwer, Phys. Rev. Lett. {\bf 95}, 195501 (2005)
\bibitem{taraskin} S. N. Taraskin and S. R. Elliott, Phys. Rev. B {\bf 59}, 8572 (1999)
\bibitem{mizu} H. Mizuno, H. Shiba, and A. Ikeda, PNAS {\bf 114}, E9767 (2017)
\bibitem{caroli} C. Caroli and A. Lemaitre, J. Chem. Phys. {\bf 153}, 144502 (2020)
\bibitem{bule} E. Bouchbinder and E. Lerner, New J. Phys. {\bf 20}, 073022 (2018)
\bibitem{asyth4} U. Buchenau, arXiv:1907.04848
\bibitem{wischi} A. Wischnewski, U. Buchenau, A. J. Dianoux, W. A. Kamitakahara, and J. L. Zarestky, Phys. Rev. B {\bf 57}, 2663 (1998)
\bibitem{schoenfeld} U. Buchenau, C. Sch\"onfeld, D. Richter, T. Kanaya, K. Kaji, and R. Wehrmann, Phys. Rev. Lett. {\bf 73}, 2344 (1994)
\bibitem{leibfried} G. Leibfried and N. Breuer, {\it Point Defects in Metals I}, Vol. 81 of Springer Tracts in Modern Physics (Springer, Berlin 1981)
\bibitem{newfrench} E. Clouet, C. Varvenne, and Th. Jourdan, Comp. Mat. Sci. {\bf 147}, 49 (2018)
\bibitem{shov} J. C. Dyre, N. B. Olsen, and T. Christensen, Phys. Rev. B {\bf 53}, 2171 (1996)
\bibitem{suck} J.-B. Suck, J. Non-Cryst. Solids {\bf 205-207}, 592 (1996)
\bibitem{scho1} A. Meyer, J. Wuttke, W. Petry, A. Peker, R. Bormann, G. Coddens, L. Kranich, O. G. Randl, and H. Schober, Phys. Rev. B {\bf 53}, 12107 (1996)
\bibitem{scho2} A. Meyer, J. Wuttke, W. Petry, O. G. Randl, and H. Schober, Phys. Rev. Lett. {\bf 80}, 4454 (1998)
\bibitem{berret} J. F. Berret and M. Meissner, Z. Phys. B - Condensed Matter {\bf 70}, 65 (1988)
\bibitem{ac} U. Buchenau, arXiv:2003.07246
\bibitem{kauz} U. Buchenau, arXiv:2101.10980
\bibitem{swap} A. Ninarello, L. Berthier, and D. Coslovich, Phys. Rev. X {\bf 7}, 021039 (2017)
\bibitem{golding} B. Golding, J. E. Graebner, A. B. Kane, and J. L. Black, Phys. Rev. Lett. {\bf 41}, 1487 (1978)
\bibitem{ray} A. K. Raychaudhuri and S. Hunklinger, Z. Physik B {\bf 57}, 113 (1984)
\bibitem{esqui} P. Esquinazi and R. K\"onig, in {\it Tunneling Systems in Amorphous and Crystalline Solids}, ed. by P. Esquinazi (Springer, Berlin 1998), p. 145 ff.
\bibitem{abs} U. Buchenau, G. D'Angelo, G. Carini, X. Liu and M. A. Ramos (unpublished).
\bibitem{sco} T. Scopigno, J.-B. Suck, R. Angelini, F. Albergamo, and G. Ruocco, Phys. Rev. Lett. {\bf 96}, 135501 (2006)
\bibitem{li} Y.Li, P. Yu, and H. Y. Bai, Appl. Phys. Lett. {\bf 86}, 231909 (2009)
\bibitem{tian} Y. Tian, Zh. Q. Li, E. Y. Jiang, Solid State Commun. {\bf 149}, 1527 (2009)
\bibitem{se} W. A. Phillips, U. Buchenau, N. N\"ucker, A. J. Dianoux, and W. Petry, Phys. Rev. Lett. {\bf 63}, 2381 (1989)
\bibitem{kob} C. Donati, J. F. Douglas, W. Kob, S. J. Plimpton, P. H. Poole, and S. C. Glotzer, Phys. Rev. Lett. {\bf 80}, 2338 (1998)
\bibitem{lind} M. L. Lind, G. Duan, and W. L. Johnson, Phys. Rev. Lett. {\bf 97}, 015501 (2006)
\end{thebibliography}
\end{document}